\documentclass[prl,twocolumn,showpacs]{revtex4}
\usepackage{bm}
\usepackage{amssymb}
\begin{document}
\title{Radiative falloff in the background of rotating black holes}
\author{Lior M.~Burko}
\affiliation{Department of Physics, University of Utah, Salt Lake
City, Utah 84112}
\author{and Gaurav Khanna}
\affiliation{Theoretical and Computational Studies Group, Southampton
College of Long Island University, Southampton, New York 11968}
\date{February 20, 2003}
\begin{abstract}
We study numerically the late-time tails of linearized fields with any
spin $s$ in the background of a spinning black hole. Our code is based on
the ingoing Kerr coordinates, which allow us to penetrate through
the event horizon. The late time tails are dominated by the
mode with the least multipole moment $\ell$ which is consistent with the
equatorial symmetry of the initial data and is equal to or greater than
the least radiative mode with $s$ and the azimuthal number $m$. 
\end{abstract}
\maketitle

The late time dynamics of black hole perturbations has been studied for
over three decades. Complete understanding of the late-time dynamics is
available for a Schwarzschild background: Generic perturbation fields of
either scalar, electromagnetic, or gravitational fields decay at late
times along an $r={\rm const}$ curve as an inverse power of
time. Specifically, linearized fields (the scalar field itself, or the
Teukolsky function $\psi$ in the gravitational case) decay as
$t^{-(2\ell+3)}$
(assuming that the initial data have compact support and are not
time-symmetric), where $\ell$ is the multipole moment of the perturbation
field \cite{price,barack,gundlach1}. This behavior was confirmed also for
fully nonlinear collapse of spherical scalar fields
\cite{gundlach2,burko-ori}. The mechanism which is responsible for this
behavior is the scattering of the field off the curvature of spacetime 
asymptotically far from the black hole. 

Because it is only the asymptotically far geometry which determines the
behavior of the late-time tails, it is natural to expect similar behavior
also when the black hole is rotating \cite{poisson}. Because spacetime is
not spherically symmetric, however, spherical-harmonic modes do not evolve
independently. Specifically, taking the initial data of the perturbation
field to be a pure $Y^{\ell m}$ mode, other modes are excited. Intuitively, 
all the
modes which are not disallowed [by symmetry requirements (such as the
equatorial symmetry of the initial data) or dynamical
considerations (such as that only modes with $-\ell\le m\le\ell$ are
allowed)] will be excited. In particular, modes with $\ell$ values
{\it smaller} than the original $\ell$ will be excited, and will dominate
at late times. (Notice that because the background is axially symmetric,
modes with different values of $m$ are not excited when
{\it linearized} perturbation theory is applied.) Accordingly, 
the late-time dynamics is dominated by the mode with the least
$\ell$ which is excited, namely the smallest $\ell$ which is not
disallowed. That is, all modes $\ell$ which are not smaller than $|m|$ and
$|s|$, where $s$ is the spin weight of the field, and
which respect the equatorial symmetry of the initial data, will be
excited. The falloff rate is then $t^{-(2\ell_{\rm min}+3)}$, where
$\ell_{\rm min}$ is the smallest mode which can be excited.

Despite the simplicity of this intuitive picture, recent papers report 
conflicting results. An analytical analysis by Hod --- in which the author
attempted to find the {\it asymptotic} behavior of the fields in the
spacetime of a Kerr black hole --- yielded results which are more complicated: 
The decay rate for a scalar field is predicted by Hod to be
\cite{hod-scalar} 
$t^{-(2\ell^*+3)}$ if $\ell^*=m$ or $\ell^*=m+1$, $t^{-(\ell^*+m+1)}$ if
$\ell^*-m\ge 2$ is even, and $t^{-(\ell^*+m+2)}$ if $\ell^*-m\ge 2$ is
odd, where $\ell^*$ is the {\it initial} value of $\ell$. For
gravitational perturbations Hod's formula is \cite{hod-prl} 
$t^{-(\ell^*+\ell_0+3-q)}$ (for axisymmetric perturbations), where
$\ell_0$ is the radiative mode with the least value of $\ell$, and $q={\rm
min}(\ell^*-\ell_0,2)$. [Different, apparently conflicting results were
reported by Barack and Ori \cite{barack-ori}. Those authors assumed that
the mode $\ell,m=0$ is present in the initial data (for $s=0$), as a
result of which it is not straightforward to confront their predictions
with Hod's.] 

Although Hod's results could be relevant for an {\it intermediate} regime
for carefully chosen parameters, they make only little sense for
describing the intended asymptotic late-time behavior. These eerie
conclusions imply that some sort of a ``memory effect'' takes place: the
field somehow ``remembers'' its initial configuration, despite being a
linearized field. We do not believe that such a ``memory effect'' is
reasonable: Take the initial data at the time $t_0$ to be those of the
pure mode $\ell^*$, such that $\ell^*$ is significantly larger than
$\ell_{\rm min}$. At the time $t_1>t_0$ the field includes, in addition to
the mode $\ell^*$, also contributions from modes $\ell<\ell^*$ because of
the excitation of other $\ell$ modes. Now the fields at $t=t_1$ can be
construed as the initial data of a new evolutionary problem. In the new
problem the initial data are a mixture of modes, such that modes $\ell$
smaller than $\ell^*$ are present \cite{poisson}. Because the mode with
the smallest existing $\ell$ value dominates at late times and determines
the decay rate of the tail, we can see no way in which the $\ell^*$ mode
can determine the asymptotic late-time tail, unless $\ell^*$ determines
which modes can and which modes cannot be excited. As in the spacetime of
a Kerr black hole it is hard to see how a scenario in which modes which
are not disallowed can still be excluded, we conclude that ``memory
effects'' are not to be expected. Hod's results, if correct,
suggest to us that an hitherto unsuspected mechanism of selection rules
inhibits the excitation of otherwise allowed modes. Such a
counter-intuitive theoretical reasoning must have strong numerical
support in order not to be discarded. 

Conclusions which apparently are similar to Hod's were obtained more
recently by Poisson \cite{poisson}, who analyzed the scalar-field tails in
a general weakly-curved, stationary, asymptotically flat spacetime. We
emphasize that unlike Hod's analysis --- which is an attempt to find the
asymptotic late-time behavior in the spacetime of a spinning black hole
--- Poisson's analysis aims at finding the behavior in a spacetime in
which curvature is weak everywhere. While Poisson's analysis and results
are correct for the spacetime he studies, one should use caution when
infering from Poisson's results on the asymptotic late-time behavior in a
Kerr geometry: Although the asymptotically-far geometries are similar, the
near-field geometries are very different. As we discuss below, that is a
crucial element in understanding the late-time behavior. 

Hod's surprising predictions agree with some reported numerical
simulations. In particular, for the case $s=0$, $\ell^*=0,m=0$ Hod's
formula predicts a decay rate of $t^{-3}$, which is indeed found
\cite{KLP96}. For the case $s=0$, $\ell^*=4,m=0$, however, Hod's formula
predicts a decay rate of $t^{-5}$, whereas the intuitive picture predicts
a decay rate of $t^{-3}$. This case was simulated numerically by Krivan
\cite{krivan}, who found a decay rate with a non-intergal index close to 
$-5.5$. Like Hod, Krivan too tried to find the asymptotic late-time
behavior in the Kerr spacetime. Some view this as a loose confirmation of
Hod's prediction \cite{poisson}, with numerical accuracy of $10\%$, and as
an invalidation of the intuitive picture. 

In this Rapid Communication we present results from independent numerical
simulations for linearized perturbation fields over a Kerr background. Our
simulations show a clear falloff rate of $t^{-3}$ for the initial data of
$s=0$, $\ell^*=4$, $m=0$. The quality of our results invalidates Hod's
prediction for the asymptotic decay rate, and points at difficulties with
Krivan's simulations or their interpretation. In all the cases we have
checked, for either a scalar or a gravitational field, we find that the
intuitive picture is correct: the late time behavior is dominated by the
mode with the lowest value of $\ell$ which can be excited. In particular,
no spooky memory effects occur.

We used the penetrating Teukolsky code (PTC) \cite{ptc}, which solves the
Teukolsky equation for linearized perturbations over a Kerr background in
the ingoing Kerr coordinates $({\tilde t},r,\theta, {\tilde \varphi})$. 
The Kerr metric is given by 
\begin{eqnarray}\label{metric}
\,ds^2=\left(1-\frac{2Mr}{\Sigma}\right)\,d{\tilde t}^2-
\left(1+\frac{2Mr}{\Sigma}\right)\,dr^2-\Sigma\,d\theta^2\nonumber \\ 
-
\sin^2\theta\left(r^2+a^2+\frac{2Ma^2r}{\Sigma}\sin^2\theta\right)
\,d{\tilde \varphi}^2-\frac{4Mr}{\Sigma}\,d{\tilde t}\,dr\nonumber \\
+
\frac{4Mra}{\Sigma}\sin^2\theta\,d{\tilde t}\,d{\tilde \varphi}+
2a\sin^2\theta\left(1+\frac{2Mr}{\Sigma}\right)\,dr\,d{\tilde \varphi}\, ,
\end{eqnarray}
where $\Sigma=r^2+a^2\cos^2\theta$, and $M,a$ are the mass and the
specific angular momentum, respectively. These coordinates are related to
the Boyer-Lindquist coordinates $(t,r,\theta,\varphi)$ through 
${\tilde \varphi}=\varphi+\int a\Delta^{-1}\,dr$ and 
${\tilde t}=t-r+r_*$, where $\Delta=r^2+a^2-2Mr$ 
and $r_*=\int(r^2+a^2)\Delta^{-1}\,dr$. Notice that ${\tilde t}$ is linear
in $t$, so that along $r={\rm const}$, $\partial /\,\partial{\tilde t}=
\partial /\,\partial t$. 

The Teukolsky equation for the function $\psi$ in the ingoing
Kerr coordinates can be obtained by implementing black hole perturbation
theory (with a minor rescaling of the Kinnersley tetrad \cite{ptc}). It is
given by 
\begin{eqnarray}\label{teukolsky}
&&
(\Sigma + 2Mr){{\partial^2 \psi}\over
{\partial \tilde t^2}} - \Delta {{\partial^2 \psi}\over
{\partial r^2}} + 2(s-1)(r - M){{\partial \psi}\over
{\partial r}} \nonumber
\\
&&
-{{1}\over {\sin \theta}}{{\partial}\over {\partial \theta}} \left (   
\sin \theta {{\partial \psi}\over {\partial \theta}}\right ) -{{1}\over
{\sin^2 \theta}}{{\partial^2 \psi}\over {\partial \tilde \varphi^2}}  
-4Mr{{\partial^2 \psi}\over {\partial \tilde t \partial r}}\nonumber\\
&&
-2a {{\partial^2 \psi}\over {\partial r\partial \tilde \varphi}}
- i {2s\cot\theta \over \sin\theta}
{{\partial \psi}\over {\partial \tilde \varphi}}
 + (s^{2}\cot^{2}\theta+s)\psi
\nonumber\\
&&
+ 2\left[{sr+ias\cos\theta+(s-1)M}\right] {{\partial \psi}\over
{\partial \tilde t}}
 = 0\, .
\end{eqnarray}

Equation (\ref{teukolsky}) has no singularities at the event horizon, and
therefore is capable of evolving data across it. The PTC implements the
numerical integration of Eq. (\ref{teukolsky}) by decomposing it into
azimuthal angular modes and evolving each such mode using a reduced 2+1
dimensional linear partial differential equation. The results obtained
from this code are independent of the choice of boundary conditions,
because the inner boundary is typically placed inside the horizon,
whereas the outer boundary is placed far enough that it has no effect on
the evolution.

The PTC has been tested in various different situations. First, it 
yields the correct complex frequencies for the quasi-normal modes of a 
Kerr black hole for a wide range of values of $a/M$. Second, it has 
also been shown to yield equivalent results in the context of the close 
limit collision of two equal mass, non-spinning, non-boosted black holes
(to ones obtained from the Zerilli formalism) \cite{gaurav}. It is stable,
and exhibits second-order convergence. 

\begin{figure}
\input epsf
\centerline{ \epsfxsize 8.0cm
\epsfbox{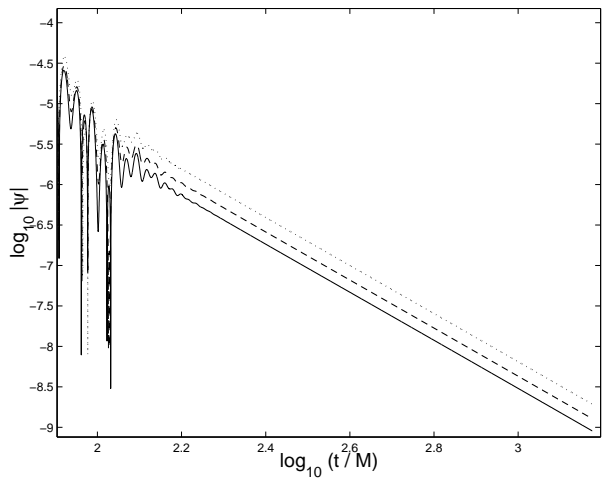}}
\caption{The Teukolsky function $\psi$ as a function of ${\tilde t}$ for
three different grid resolutions, for $s=0$, $a/M=0.9$,
$\ell^*=4$, and $m=0$. Dotted line: 8000 steps in $r$ and $40$
steps in $\theta$. Dashed line: 10000 steps in $r$ and $50$
steps in $\theta$. Solid line: 12000 steps in $r$ and $60$
steps in $\theta$. The time step is always taken to be half the step in
$r$. The data are shown along $r=20M$ on the equatorial plane.} 
\label{fig1}
\end{figure}

We next set $a/M=0.9$, $s=0$, and $\ell^*=4$, $m=0$. The
initial gaussian perturbation is taken to be a mixture of 
ingoing and outgoing waves, and centered about $r=20M$ with a width of
$4M$. As discussed above, our
expectations are that all the even $\ell$ modes are excited (respecting
the equatorial symmetry of the initial data). The least $\ell$ mode which
is excited is the $\ell=0$ mode, so that the decay rate we expect is
$t^{-3}$. In contrast, the prediction of Hod is for a
decay rate of $t^{-5}$. Figure \ref{fig1} shows the Teukolsky function
$\psi$ for these initial data for $\theta=\pi/2$ (the equatorial
plane) for three different resolutions. The data clearly indicate
stability and second-order convergence. 

\begin{figure}
\input epsf
\centerline{ \epsfxsize 8.0cm
\epsfbox{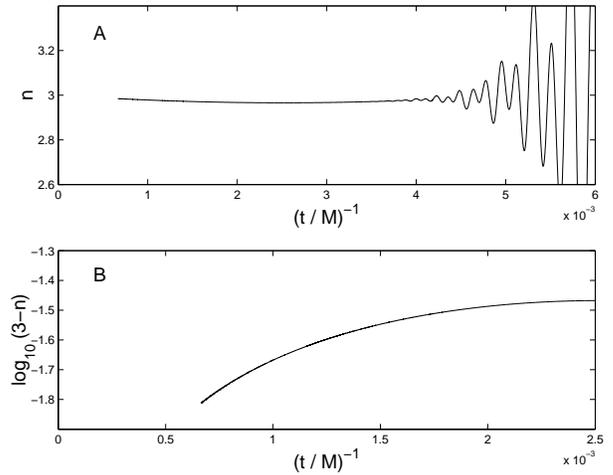}}
\caption{The local power index for the same data as in Fig.~\ref{fig1}.
Upper panel (A): $n$ as a function of $M/{\tilde t}$. Lower panel (B): 
$\log_{10} |3-n|$ as a function of $M/{\tilde t}$. The data are shown for 
an equatorial curve at $r=20M$.}
\label{fig2}
\end{figure}

\begin{figure}
\input epsf
\centerline{ \epsfxsize 8.0cm
\epsfbox{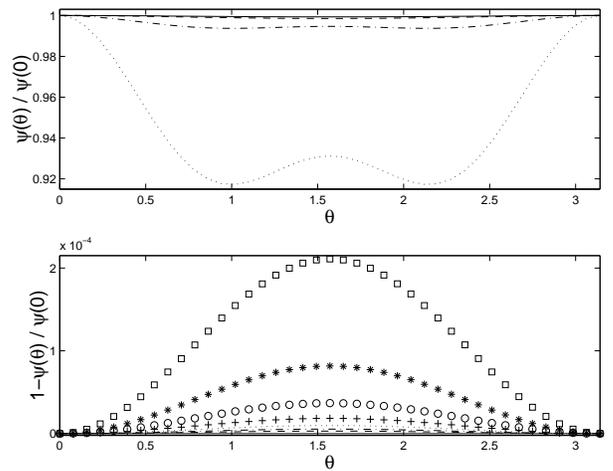}}
\caption{The normalized Teukolsky function $\psi / \psi(\theta=0)$ as a
function of $\theta$ at $r=25M$, for the same data as in Fig.~\ref{fig1},
for various values of ${\tilde t}$. Upper panel: At ${\tilde t}=150M$
(dotted line), $200M$ (dash-dotted), $250M$ (dashed), and $300M$ (solid
line). Lower panel: At ${\tilde t}=400M$ ($\square$), $500M$ ($\ast$),
$600M$  ($\circ$), $700M$ (+), $800M$ (dotted line), $900M$ (dash-dotted),
$1000M$ (dashed), and $1500M$ (solid line).}
\label{fig3}
\end{figure}

A decay rate of about $t^{-3}$ is already clear from
Fig.~\ref{fig1}. Evaluating the decay rate from the slope of the field is
very inaccurate: The slope then depends on the interval one chooses, and
also on the presence of subdominant modes. The first difficulty can be
handled by considering the {\it local power index} 
$n$ \cite{burko-ori}, which we define as $n\equiv -({\tilde
t}/\psi) \,\partial_{\tilde t}\psi$. 
The second difficulty can be handled by extrapolating $n$ to timelike
infinity.  
Figure \ref{fig2}A shows $n$ as a function of $M/{\tilde t}$. Timelike
infinity is at zero, and both the regime where the field is dominated by
the quasi-normal ringing and the regime where the field is dominated by
the power-law tails are shown. The local power index $n=2.9846$ at
${\tilde t}=1500M$. Figure \ref{fig2}B shows the behavior of $3-n$ as a
function of $M/{\tilde t}$. Clearly, $n$ gets closer with time to the
expected value of $3$. In fact, extrapolating $n$ to ${\tilde t}\to\infty$
using Richardson's deferred approach to the limit, we find the asymptotic
value of $n$ to be $n_{\infty}=3.0003\pm 0.0011$. Our results suggest that
the late-time field is dominated by the $\ell=0$ mode. We checked this by
plotting $\psi$ as a function of $\theta$ in
Fig.~\ref{fig3} for different values of ${\tilde t}$. We indeed find that
$\psi$ quickly loses any dependence on $\theta$, such that at late times 
it is indeed described by the $\ell=0$ mode. Any dependence of $\psi$
on $\theta$ is smaller than 3 parts in $10^6$ at ${\tilde t}=1000M$. 

\begin{figure}
\input epsf
\centerline{ \epsfxsize 8.0cm
\epsfbox{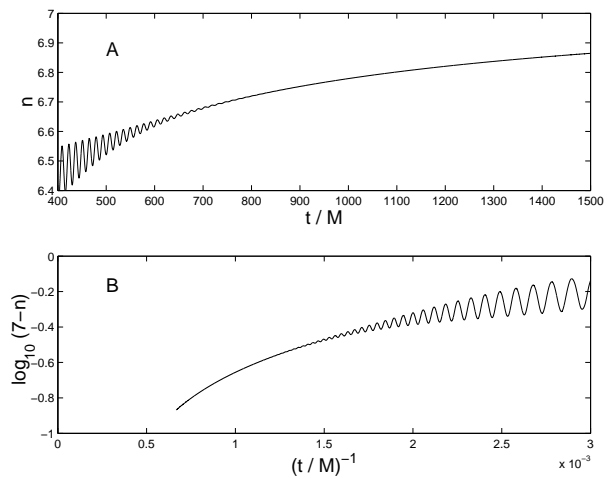}}
\caption{The local power index for $s=2$, $a/M=0.3$, $l^*=6$, and $m=0$. 
Upper panel (A): $n$ as a function of ${\tilde t}/M$. Lower panel (B): 
$\log_{10}(7-n)$ as a function of $M/{\tilde t}$. The data are taken for
an equatorial curve at $r=20M$.}
\label{fig4}
\end{figure}

Next, we present results for the behavior of fields with higher
spins. We set the parameters to $s=2$, $a/M=0.3$, and initial $l^*=6$,
$m=0$. The pulse is again centered about $r=20M$ with a width of $4M$. The
prediction of Hod's formula for this case is a decay rate of $t^{-9}$. In
this case our expectations are that the least $\ell$ mode to be excited is
the $\ell=2$ mode. Consequently, we expect the decay rate to be
$t^{-7}$. This is indeed confirmed in Fig.~\ref{fig4}A, which shows the
local power index $n$ as a function of ${\tilde t}/M$, and in
Fig.~\ref{fig4}B which displays $7-n$ as a function of $M/{\tilde t}$. 
At ${\tilde t}=1500M$, we find that $n=6.8646$. Extrapolating $n$ to
timelike infinity, we find that $n_{\infty}=7.01\pm 0.03$, in agreement 
with our expectations. 

Our results clearly show that starting with a pure mode $\ell^*,m$, the
late-time decay rate is dominated by the least mode $\ell_{\rm min}$
which is consistent with the equatorial symmetry of the initial data and
is equal to or greater than the least radiative mode $\ell_0={\rm
max}(|s|,|m|)$. The late-time decay rate is given by $t^{-(2\ell_{\rm
min}+3)}$. Our conclusions are in sharp disagreement with the recent
predictions by Hod \cite{hod-prl,hod-scalar}. Hod's
analysis is in the frequency domain, and carried to leading order in
$\omega$, the angular frequency. That approach is very successful in the
background of a Schwarzschild black hole, where it reproduces the known
results \cite{andersson}. The understanding that the power-law tails
result from scattering of the field at asymptotically large distances
implies that it is only the small $\omega$ which are responsible for the
tails. That is indeed the case with a Schwarzschild black hole. We
conjecture that it would also be the case for a Kerr black hole, if there
were no excitations of dominating modes which are not present in the
initial data. For example, in the case of a scalar field with
$\ell^*=m=0$, the dominating mode is already present in the initial
data. Considering only the small $\omega$  contributions 
indeed produces a result in agreement with numerical simulations. When the
dominating mode is not present in the initial data, however, it needs
first to be excited. If it is excited (with any nonzero amplitude), the
small $\omega$ approximation may produce the correct result for the decay
rate. However, mode-excitation is an effect which is nonlinear in the
gravitational potentials, and is strongest in the near zone. This suggests
to us that a leading order (in $\omega$) analysis will not, in general, get all
the excited modes right. It might be the case that higher orders in
$\omega$ are necessary in order to get all the modes which are
excited. Our numerical results indeed show, that when the least mode which 
can be excited $\ell_{\rm min}$ is ``far'' from the initial $\ell^*$,
that technique does not produce the former: For example, for initial
$\ell^*=4$ and $m=0$, the leading order in $\omega$ analysis was able to
get the $\ell=1$ mode excited (as is manifested by Hod's decay rate of
$t^{-5}$), but not the $\ell=\ell_{\rm min}=0$ mode (which implies a decay
rate of $t^{-3}$). We suggest, that although a frequency-domain analysis
is capable of getting the decay rate right, it should include an expansion
to higher orders in $\omega$. Such an expansion would be a formidable
endeavor. In a similar way, by taking spacetime to be
weakly curved everywhere, Poisson tacitly assumed that it is just the
far-zone part of the field which is important. (In Poisson's case, we
emphasize, this assumption is well justified, because in the spacetime
studied by Poisson spacetime is nowhere strongly
curved. Incidentally, Poisson suggests a selection-rule mechanism in the
spacetime he studied, which is related to the remarkable vanishing of
terms in the initial data in the transformation from spheroidal to
spherical coordinates. The mechanism suggested by Poisson demonstrates how
indeed Hod's results could be correct in that context. However, no such
mechanism is offered for a Kerr spacetime.) That assumption is equivalent
to taking the large-$r$ approximation, or the small-$\omega$
approximation. Consequently, Poisson and Hod make, in fact, the same kind
of approximation, such that it is not surprising that they obtain the same
results. We emphasize, that Poisson acknowledges that effects which are
nonlinear in the gravitational potentials may produce modes with $\ell$
values which are smaller than those obtained by him. Poisson then remarks,
that no such effects have been reported on in the literature. Evidence for
such an effect is precisely what we find here. Although the late-time
expansion method \cite{barack-ori} does not seem to suffer from similar
weaknesses, it is hard to apply for the problem of interest. Starting with
an initial $\ell^*$ which is ``far'' from the least mode $\ell_{\rm min}$
to be excited, the method of Ref.~\cite{barack-ori} requires many
iterations in order to find the excited mode $\ell_{\rm
min}$. Specifically, three iterations are required
in order to find the $\ell=0$ mode starting with $\ell^*=4,m=0$. Carrying
this iterative scheme in practice seems like a daunting task. We would
like to repeat, that while Hod's method fails to obtain the correct {\it
asymptotic} decay rate, it may still be useful in determining an {\it
intermediate} behavior for carefully chosen parameters. 

Lastly, our results are in disagreement also with the numerical results of
Krivan \cite{krivan}, who reported on a fractional power-law index which
is about $-5.5$ for the case of initial $s=0$, $\ell^*=4$ and
$m=0$. While we cannot point with certainty to the reason why Krivan's
simulations produce a result for the asymptotic late-time behavior 
which is at odds with ours, we would like to
mention some of the factors which may be responsible: Krivan takes the
black hole to spin exceedingly fast. In fact, Krivan takes
$a/M=0.9999$. The high spin of the black hole may act in two ways: First,
it slows down the decay rate of the quasi-normal ringing, such that longer
integration times are required in order to obtain the tails. Second, the
numerical solution of the Teukolsky equation is more sensitive and harder
when the spin is very high. Another factor is related to the location and
the direction of Krivan's initial perturbation. Krivan takes the
perturbation to be centered around $r_*/M=100$, and to have a very large
width (of $100M$). Also, the perturbation is purely outgoing on
the initial slice. We thus conjecture that the dominating $\ell=0$ mode is
excited only with a very low amplitude, because most of the perturbation
field does not probe the strong-field region. This, in addition to the
great distance and width of the initial perturbation, may combine into 
late-time tails whose asymptotic behavior becomes evident only at very
late times, to which Krivan's simulations have not arrived. 

The picture which arises for linearized perturbations in the background
of a spinning black hole is simpler than that which is implied by. 
However, we expect the picture to be even simpler than that for
fully nonlinear perturbations: When the initial perturbation is not
axially symmetric, the evolving spacetime will not be axially symmetric
either. Consequently,
the $m$ value of the field will not be conserved, and different values of
$m$ will also be excited, preserving only the equatorial symmetry of the
initial data. We therefore expect a fully nonlinear evolution to yield
results which are simpler than those obtained from a linearized
analysis: Because $m$ is no longer fixed, the restriction of $\ell_0$ is
no longer so strict: $\ell_0=|s|$, and the dominating mode is
simply the least $\ell$ mode which is consistent with the equatorial
symmetry which is equal to or greater than $\ell_0$. We thus
expect generic tails to always have a decay rate of $t^{-(2|s|+3)}$. The
more complicated results of this Rapid Communication then are an artifact
of the linearization: the full theory is simpler.

We thank Eric Poisson and Richard Price for discussions. This research was
supported by NSF grants PHY-9734871 and PHY-0140236. Initial work on this
research was done while LMB was at the California Institute of Technology,
where it was supported by NSF grant PHY-0099568. We thank the Center for
Gravitational Physics and Geometry at Penn State for computational
facilities.

\end{document}